\documentclass[a4paper,10pt]{article}

\usepackage[centertags]{amsmath}
\usepackage{amsfonts}
\usepackage{amssymb}
\usepackage{amsthm}
\usepackage{newlfont}

\theoremstyle{plain}
  \newtheorem{theorem}{Theorem}[section]
  
  \newtheorem{proposition}[theorem]{Proposition}
  
\theoremstyle{definition}

\theoremstyle{remark}

\numberwithin{equation}{section}

  \let\de=\delta \let\ep=\epsilon
   
\let\ka=\kappa

   \let\Om=\Omega

\newcommand{\caF}{{\mathcal F}}

\newcommand{\bbN}{{\mathbb N}}

\newcommand{\bbR}{{\mathbb R}}

\newcommand{\opunit}{\text{1}\kern-0.22em\text{l}}

\newcommand{\un}[1]{\underline{#1}}

\newcommand{\rel}{\,|\,}

\begin{document}

\begin{center}
\noindent{\large \bf  \rm{H}-Theorems from Autonomous Equations} \\

\vspace{15pt}

{\bf Wojciech De Roeck}\footnote{Aspirant FWO, U.Antwerpen} , {\bf
Christian Maes}\footnote{email: {\tt
christian.maes@fys.kuleuven.be}} \\
Instituut voor Theoretische Fysica, K.U.Leuven
(Belgium)\\\vspace{10pt}
 {\bf Karel Neto\v{c}n\'{y}\footnote{K.N. acknowledges the support
from the project AVOZ10100520 in the Academy  of Sciences of the
Czech Republic.}}\\
 Institute of Physics AS CR, Prague
\end{center}

\vspace{20pt} \footnotesize \noindent {\bf Abstract: } The
\rm{H}-theorem is an extension of the Second Law to a
time-sequence of states that need not be equilibrium ones.  In
this paper we review and we rigorously establish the connection
with macroscopic
autonomy.\\
If for a Hamiltonian dynamics for many particles, the macrostate
evolves autonomously, then its entropy is non-decreasing as a
consequence of Liouville's theorem. That observation, made since
long, is here rigorously analyzed with special care to reconcile
the application of Liouville's theorem (for a finite number of
particles) with the condition of autonomous macroscopic evolution
(sharp only in the limit of infinite scale separation); and to
evaluate the presumed necessity of a semigroup property for the
macroscopic evolution.

\vspace{5pt}

 \vspace{20pt} \normalsize

\section{Introduction}
\bibliographystyle{plain}
The point of the present paper is to make mathematically precise
the application of Liouville's theorem in microscopic versions or
derivations of the Second Law, under the assumption that an
autonomous evolution is verified for the macroscopic variables in
question.  Microscopic versions of the Second Law, or perhaps more
correctly, generalizations of the Second Law to nonequilibrium
situations, are here referred to as \rm{H}-Theorems. \\

The stability of points of a dynamical system can be demonstrated
with the help of Lyapunov functions.  Yet in general these
functions are hard to find
--- there does not exist a construction or a general algorithm to obtain
them.   On the other hand, when the differential equation has a
natural interpretation, as with a specific physical origin, we can
hope to improve on trial and error. Think of the equations of
irreversible thermodynamics where some approach to equilibrium is
visible or at least expected. Take for example the diffusion
equation
\begin{equation}\label{dif}
\frac{\partial n_t(r)}{\partial t} + \nabla\cdot J_r(n_t) = 0
\end{equation}
for the particle density $n_t(r)$ at time $t$ and at location $r$
in some closed box. That conservation equation is determined by
the current $J_r$ depending on the particle density via the usual
phenomenology
\begin{eqnarray}\nonumber
J_r(n_t) &=& \frac 1{2}\chi(n_t(r))\,\nabla s'(n_t(r))\\\nonumber
& = & -\frac 1{2} D(n_t(r))\,\nabla n_t(r)
\end{eqnarray}
Here $D(n_t(r))$ is the diffusion matrix, connected with the
mobility matrix $\chi$ via
\[
\chi(n_t(r))^{-1} D(n_t(r)) = -s''(n_t(r)) \mbox{ Id}
\]
 for the
identity matrix Id and the local thermodynamic entropy $s$. From
irreversible thermodynamics the entropy should be monotone and
indeed it is easy to check that $\int dr \, s(n_t(r))$ is
non-decreasing in
time $t$ along \eqref{dif}.\\

Such equations and the identifications of monotone quantities are
of course very important in relaxation problems.  A generic
relaxation equation is that of Ginzburg-Landau.  There an order
parameter $m$ is carried to its equilibrium value via \[
\frac{dm}{dt} = -D \frac{\delta\Phi}{\delta m} \] where $D$ is
some positive-definite operator, implying
\[
 \frac{d\Phi}{dt} \leq
0 \] for $\Phi(m)$ for example the Helmholtz free energy.\\
 That
scenario can be generalized. We are given a first order equation
of the form
\begin{equation}\label{dif1}
  \frac{dm_{t}}{dt}=F(m_t),\quad m_t\in \bbR^{\nu}
\end{equation}
with solution  $m_t = \phi_{t}(m)$.   Yet it is helpful to imagine
extra ``microscopic'' structure. One supposes that \eqref{dif1}
results from a law of large numbers in which $m_t$ is the
macroscopic value at time $t$ and $\phi_t$ gives its autonomous
evolution.  At the same time, there is an entropy $H(m_t)$
associated to the macroscopic variable and one hopes to prove that
$H(m_t) \geq H(m_s)$ for $t\geq s$.   That will be explained and
mathematically detailed starting with
Section \ref{classet}.\\

Usually however, from the point of view of statistical mechanics,
the problem is posed in the opposite sense. Here one looks for
microscopic versions and derivations of the Second Law of
thermodynamics. One starts from a microscopic dynamics and one
attempts to identify a real quantity that increases along a large
fraction of trajectories. We will show that such an \rm{H}-theorem
is valid for the Boltzmann entropy when it is defined in terms of
these macroscopic observables that satisfy an autonomous equation
(Propositions \ref{ja1} and \ref{ja5}).  In that context, we also
discuss the role of the semigroup property of the macroscopic
evolution, the influence of reversibility in the microscopic
dynamics and the relation with conditions of propagation of
constrained equililibrium. These results are to be considered as
mathematical precisions of what has been known by many for a long
time.  In particular, our motivation does not come from something
physically problematic in the derivation of the Second Law. The
main consideration in problems of relaxation to equilibrium is the
enormous scale separation between the microscopic and the
macroscopic worlds: the volume in phase space that corresponds to
the equilibrium values of the macroscopic quantities is so very
much larger than the volume of nonequilibrium parts.  A
thermodynamic entropy difference $s' - s$ per particle of the
order of Boltzmann's constant $k_B$ is physically reasonable and
corresponds to a total reversible heat exchange $T(S'-S)=NT(s'-s)$
of about 0.5 Joule at room temperature $T$ and to a phase volume
ratio of
\[
\frac{W}{W'} = \exp - (S' -S)/k_B = e^{-10^{20}}
\]
when the number of particles $N=10^{20}$.\\

\section{Heuristics of an H-theorem and main questions}\label{Htheorem}

\subsection{Heuristics}
The word \rm{H}-theorem originates from Boltzmann's work on the
Boltzmann equation and the identification of the so called
\rm{H}-functional.  The latter plays the role of entropy for a
dilute gas and is monotone along the solution of the Boltzmann
equation. One often does not distinguish between the Second Law
and the \rm{H}-theorem. Here we do (and the entropy will from now
on be
denoted by the symbol $H$).\\

The heuristics is simple: consider here an autonomous
deterministic evolution taking macrostate $M_s$ at time $s$ to
macrostate $M_t$ at time $t\geq s$.  Then, under the Liouville
dynamics $U$ the volume in phase space $|M_s| = |U M_s|$ is
preserved.  On the other hand, since about every microstate $x$ of
$M_s$ evolves under $U$ into one corresponding to $M_t$, we must
have, with negligible error that $UM_s \subset M_t$.  We conclude
that $|M_s| \leq |M_t|$ which gives monotonicity of the Boltzmann
entropy $H= k_B \log |M|$.\\
That key-remark has been made before, most clearly for the first
time on page 84 in \cite{jay3}, but see also e.g. \cite{jay1} page
9--10, \cite{jay2} page 280--281, \cite{fig} Fig.1 page 47,
\cite{ref} page 278, page 301, and most recently in
\cite{gl,ggl}.\\
 The set-up we start from in the next section is a
classical dynamical system and we show in what sense one can say
that when a collection of variables obtains an autonomous
evolution, the corresponding entropy will be monotone.  A somewhat
introductory but rather formal and abstract argument goes as
follows.\\

Consider a transformation $f$ on states $x$ of a measure space
$(\Omega, \rho)$. The measure $\rho$ is left invariant by $f$.
Suppose there is a sequence $(M_n), n \in \bbN$ of subsets $M_n
\subset \Omega$ for which
\begin{equation}\label{rep}
\rho((f^{-1}M_{n+1})^c \cap M_n) =0
\end{equation}
In other words, $M_{n+1}$ should contain about all of the image $f
M_n$. Then,
\[
\rho(M_{n+1}) = \rho(f^{-1}M_{n+1}) \geq \rho(f^{-1}M_{n+1} \cap
M_n) = \rho(M_n)
\]
and $\rho(M_n)$ or $\log \rho(M_n)$ is a non-decreasing sequence.\\

One can think of the $M_n$ above as macrostates that are
successively visited by the microstate in the course of time.  One
has in mind a partition of $\Omega$ and a map $M$ that sends each
$x\in \Omega$ to the member set of the partition to which it
belongs. The partition corresponds to dividing the phase space
according to the values of the relevant macroscopic variables. The
entropy can be defined on microstates as
\[
H(y) \equiv \log \rho(M (y)),\quad y\in \Omega
\]
and if for some $x\in \Omega$ the sequence $M_n = M(f^nx)$
satisfies \eqref{rep}, then the entropy
\begin{equation}\label{se} H(x_n)\equiv H(f^n(x)) =
\log \rho(M(f^n(x)))
\end{equation}
 is
monotone along the path starting from $x$. The condition
\eqref{rep} basically requires that  the transformation $f$ gets
replaced on  the level of the sets $M_n$, i.e., on the macroscopic
level, with a new
autonomous dynamics.\\

\subsection{Questions}\label{que}
The previous heuristics, be it verbal or in a more abstract
notation, calls for some further questions and warnings. Below
follows our motivation to add more mathematical precision in the
next Section \ref{classet}. The point is simply that the
heuristics above cannot be taken too literally; otherwise it boils
down to trivialities. The autonomy has to be relaxed to allow for
nontrivial statements, and as a consequence also the semigroup
structure of the  macroscopic dynamics (essential for the argument
and trivially true in \eqref{rep}) might be lost in general and
must be enforced.

\subsubsection{Scale separation}
Remark that Liouville's theorem (or the invariance of the natural
measure) is essential in the above heuristics. It employs a finite
number of particles.  Yet, the autonomy of the macroscopic
equation is probably only satisfied in some hydrodynamic limit
where also the number of particles goes to infinity. In
particular, \eqref{rep} is only expected  verified in some limit
where the degrees of freedom $N\uparrow \infty$. There is thus the
question, how to mathematically formulate the conditions and
conclusions in a way that is physically reasonable. In fact, if
\eqref{rep} were satisfied \emph{exactly} for a finite system, the
H-function would necessarily be constant, as we will now show.

We take the same start as above but we assume also that f is
bijective and that there is a countable partition $(P_i), i\in I $
of $\Omega$ with $ \rho(P_{i}) >0, i\in I$, for $\rho$ a
probability measure that is left invariant by $f$.  We assume
autonomy in the sense that there is a map $\phi$ on $I$ with $f
P_i \subset P_{\phi i}$ up to $\rho-$measure zero, i.e., $\rho(
(f^{-1}P_{\phi i})^c \cap P_i)=0$, cf.~\eqref{rep}. We show that
under these conditions the entropy is constant, i.e.,
\[
\rho(P_i)=\rho(P_{\phi i})  \textrm{ for all } i
\]
To see that, note that for all $n \in \bbN$, $\rho(P_{\phi^n i})
\geq  \rho(P_{i})$. Therefore, as $\rho$ is normalized, there must
be $n <m$ such that $\phi^n i=\phi^m i$. On the other hand, always
$\rho-$measure zero, both
$fP_{\phi^{n-1}i},fP_{\phi^{m-1}i}\subset
P_{\phi^{m}i}=P_{\phi^{n}i}$.  If now $\phi^{n-1}i \neq \phi^{m-1}
i$, we have the contradiction
\begin{eqnarray}
\rho(P_{\phi^m i}) &\geq & \rho(P_{\phi^{n-1}i})  +
\rho(P_{\phi^{m-1}i}) \nonumber\\&\geq& \rho(P_{\phi^{n-1} i})+
\rho(P_{\phi^{n} i}) = \rho(P_{\phi^{n-1} i})+ \rho(P_{\phi^{m}
i})\nonumber\\&
>& \rho(P_{\phi^{m} i})
\end{eqnarray}
Hence the trajectory $\phi^n i, n \in \bbN$ is a closed cycle,
$\phi^ki=i$ for some $k$.  As a consequence, the entropy is
strictly constant. \\

The above scenario corresponds to assuming full autonomy for the
macroscopic dynamics for a finite system.  As shown, then, the
macroscopic dynamics cannot be irreversible.   On the other hand,
in a thermodynamic ``infinite'' system, it is typically impossible
to find a partition into macrostates such that each macrostate has
a non-zero measure; the equilibrium state will have full measure.
Hence it is necessary to add more structure and the
autonomy should only hold in a thermodynamic limit.\\

All that does not mean that we need to look ``in the thermodynamic
limit''. In fact, physically speaking, we are very much interested
in a statement which is true for a very large but finite number of
degrees of freedom (as also stressed in section 2.2.3). Such a
statement will come in the Section \ref{findim}.

\subsubsection{Semigroup property}
Dynamical equations for macroscopic degrees of freedom have
varying mathematical properties.  Sometimes the macroscopic
dynamics is explicitly time-dependent, such as in the case when
some external force is present, and sometimes the evolution is
given via a differential equation of higher order or perhaps via
an integro-differential equation, containing physically important
so called memory terms, or the macroscopic dynamics could very
well be not deterministic at all. It is therefore appropriate to
be more explicit about what we mean by autonomy and under what
mathematical conditions a standard H-theorem results.  The
conclusion in general will be that the H-theorem with the
Boltzmann entropy as presented below is verified when the
autonomous equation is in a sense of first order.

 To see what can
happen otherwise, let us imagine that a macroscopic degree of
freedom $m_t$ in a thermodynamic system evolves according to
  $m_t = m_0 r^t\,\cos \omega
t, |r|<1$.  An example can be found in Section 3.3 of \cite{qKac}
--- it is like the position of a pendulum, swinging with decreasing amplitude
around its equilibrium.  Obviously, if we consider $m_t$ as the only macrovariable, then the
H-function only depends on $m_t$ and hence it cannot be monotone.
Nevertheless the equation for $m_t$
is completely determined by the value $m_0$ at time zero. It is
only after adding other degrees of freedom, like the speed of change of the degree of freedom $m_t$, that we get a monotone H-function. The choice of
macroscopic variables is therefore absolutely relevant. It will
decide whether the macroscopic values at a later time are
determined from the value of the macroscopic variables at any
earlier time, as required in \eqref{rep}. If not, one can imagine
a macroscopic dynamics satisfying only
\begin{equation}\label{nrep}
  \rho((f^{-n} M_n)^c \cap M_0) = 0 \qquad n \in \bbN
\end{equation}
i.e., the initial macrostate $M_0$ determines the whole trajectory
as well but the macrodynamics is possibly not satisfying
\eqref{rep}. Loosely speaking, that can happen when almost all of
a macrostate $M_1$ is mapped into macrostate $M_2$, and nearly all
of $M_2$ is mapped into $M_3$, but  $fM_1 \subset M_2$ is not
typically mapped into $M_3$.\\
To get an \rm{H}-theorem for the
more general situation (e.g. for higher order differential
equations), the entropy \eqref{se} would have to be generalized.
In other words, condition \eqref{rep} of autonomy goes hand in
hand with
the interpretation of \eqref{se} as an entropy.\\

At the other side, one wants to see whether conditions referred at
as repeated randomization or molecular chaos which are indeed
sometimes directly used in justifications of Markov approximations
or in the derivation of autonomous macroscopic equations, are
necessary for an H-theorem.  Microscopically speaking, these
conditions not only demand that $fM_n \subset M_{n+1}$, as in
\eqref{rep}, but they also ask that $fM_n$ is so to speak randomly
distributed in $M_{n+1}$.  It is as if at every time $n$, the
microscopic state can be thought of as randomly drawn from the set
$M_n$.  We will show in Section \ref{propa} how such a type of
chaoticity assumption is stronger than what we effectively mean by
autonomy.

\subsubsection{Corrections to the H-theorem}

For a system composed of many particles, we can expect a (first
order) autonomous evolution over a certain time-scale for {\it a
good choice} of macroscopic variables.   In that case \eqref{se}
coincides with the Boltzmann entropy: it calculates the volume in
phase space compatible with some macroscopic constraint (like
fixing energy and some density- or velocity profile).  The
identification with the thermodynamic entropy (in equilibrium)
then arises from considerations of equivalence of ensembles. In a
way, the \rm{H}-theorem is a nonequilibrium version of the Second
Law --- not only considering initial and final equilibria but also
the entropy of the system as it evolves possibly away from
equilibrium.\\
Two related questions enter then.  Whether one can see in what
sense the H-theorem is approached as the number of degrees of
freedom $N\uparrow +\infty$, and whether the corresponding
``finite world" H-function can be seen to be monotone along the
microscopic trajectory.  The H-function then needs to be defined
on the microscopic state and to be followed as the microscopic
dynamics prescribes. That will be done in Section \ref{findim}\\

All of what follows concentrates on mathematically precise and
physically reasonable formulations of \eqref{rep} and \eqref{se}
to obtain monotonicity of entropy.
 The main purpose is therefore to
clarify a theoretical/mathematical question; not to include new
results for specific models. The only difficulty is to identify
the appropriate set of assumptions and definitions; from these the
mathematical
 arguments will be relatively short and easy.

\section{Classical dynamical systems}\label{classet}
Let $N$ be an integer, to be thought of as the number of degrees
of freedom or as a scaling parameter, that indexes the dynamical
system $(\Omega^N,U^N_t,\rho^N)$.  $\Omega^N $ is the phase space
with states $x \in \Omega^N$ and is equipped with a probability
measure $\rho^N$, invariant under the dynamics
$U^N_t:\Omega^N \rightarrow \Omega^N$.\\

We suppose a map
\begin{equation}\label{macromap}
m^N : \Omega^N \rightarrow \cal{F}
\end{equation}
which maps every state $x$ into an element $m^N(x)$ of a metric
space $(\cal{F},d)$ (independent of $N$).
  For $\cal{F}$ one can have in mind
$\bbR^n$ for some integer $n$ or a space of real-valued functions
on a subset of $\bbR^n$, with the interpretation that $m^N(x)$
gives the macroscopic state
 corresponding to the microscopic state $x$.
 For $m,m' \in \cal{F}$ and $\delta >0$ we introduce the
 notation $m'\stackrel{\de}{=}m$ for $d(m',m) \leq \delta$.

\subsection{Infinite scale separation}\label{inflim}

We start here by considering the limit $N\uparrow+\infty$. In that
limit the law of large numbers starts to play with deviations
governed by
\begin{equation}\label{conbe}
H(m)\equiv\lim_{\delta \downarrow 0}\lim_{N \uparrow +\infty}
\frac{1}{N}\log{\rho^N \big(m^N(x) \stackrel{\de}{=} m
\big)},\quad m\in \cal{F}
\end{equation}
That  need not exist in general, but we make that definition part
of our assumptions and set-up. For what follows under Proposition
\ref{ja1} it is in fact sufficient to take the lim sup in
\eqref{conbe} (if we also take the lim sup in the next
\eqref{markov1}) but for simplicity we prefer here to stick to the
full limit.  The limit \eqref{conbe} is then a natural notion of
 entropy \`a la Boltzmann; see later in \eqref{finite entropy} and below for a ``finite'' version.\\

The macroscopic observables are well-chosen when they satisfy an
autonomous dynamics, sharply so in the proper limit of scales.
Here we assume dynamical autonomy in the following rather weak
sense: there is an interval $[0,T]$ and a map  $\phi_t :\cal{F}
\rightarrow \cal{F}$ for all $t\in [0,T]$ such that $\forall m \in
\cal{F}, \forall \delta >0 $, and $0\leq s\leq t\leq T$
\begin{equation} \label{markov1}
\lim_{\ep \downarrow 0}\lim_{N \uparrow +\infty}
\frac{1}{N}\log\,\rho_N \Big(m^N(U^N_t x) \stackrel{\de}{=}
\phi_t(m) \big| m^N(U_s^N x) \stackrel{\ep}{=} \phi_s(m) \Big) =0
\end{equation}
\begin{proposition}\label{ja1}
$\forall m \in \cal{F}$ and for all $0\leq s\leq t\leq T$,
\begin{equation}\label{monot}
 H(\phi_t(m)) \geq H(\phi_s(m))
\end{equation}
\end{proposition}
\begin{proof}
Writing out $H(\phi_t(m))$ we find that for every $\ep >0$
\begin{eqnarray}\label{bounds}
\log{\rho^N \big(m^N( x) \stackrel{\de}{=} \phi_t(m) \big)} &=&
\log{\rho^N \big(m^N( U^N_t x) \stackrel{\de}{=} \phi_t(m) \big)}\nonumber\\
\geq && \log{\rho^N \Big(m^N( U^N_t x) \stackrel{\de}{=} \phi_t(m)
\big|m^N( U_s^N x) \stackrel{\ep}{=} \phi_s(m) \Big)}\nonumber\\
+ &&\log \rho^N\big(m^N( U_s^N x) \stackrel{\ep}{=} \phi_s(m)\big)
\end{eqnarray}
The equality uses the invariance of $\rho^N$ and we can use that
again for the last term in \eqref{bounds}.  We divide
\eqref{bounds} by $N$ and we first take the limit $N\uparrow
+\infty$ after which we send $\ep \downarrow 0$ and then
$\delta\downarrow 0$.
\end{proof}
Condition \eqref{markov1} is much less than requiring a strict
macroscopic autonomy.  We do for example not suppose that the
macroscopic trajectory is uniquely determined.  In fact, condition
\eqref{markov1} is consistent with a large class of stochastic
macroscopic dynamics too.\\
A sufficient condition for \eqref{markov1} will follow in Section
\ref{propa}.\\
The case of a stochastic microscopic dynamics will be addressed in
Section \ref{adrem}.

\subsubsection{Semigroup property} If the dynamics $(U^N_t)$
satisfies the semigroup property
\begin{equation}\label{semi}
U^N_{t+s}=U^N_{t}U^N_{s} \qquad t,s \geq 0
\end{equation}
and there is a unique macroscopic trajectory $(\phi_t)$ satisfying
\eqref{markov1}, then
\begin{equation}\label{memi}
\phi_{t+s}=\phi_{t} \circ \phi_{s} \qquad t,s \geq 0
\end{equation}
In practice the map $\phi_{t}$ will mostly be the solution of a
set of first order differential equations.\\
Observe then that \eqref{memi} combined with \eqref{markov1} for
$s=0$ also yields the full monotonicity \eqref{monot}.

\subsubsection{Reversibility}
Equation \eqref{markov1} invites the more general definition of a
large deviation rate function for the transition probabilities
\begin{eqnarray}\label{ldrf}
&&-J_{t,s}(m,m') \equiv  \\\nonumber&&\lim_{\de\to 0} \lim_{\ka
\to 0} \lim_{N \to \infty} \frac{1}{N}
  \log \rho^N(m^N(U^N_t x) \stackrel{\de}{=} m'\rel  m^N(U^N_s x) \stackrel{\ka}{=} m  ), \qquad t \geq s
\end{eqnarray}
which we assume exists.  The bounds of \eqref{bounds} give
\begin{equation}\label{bb}
  H(m') \geq H(m) - J_{t,s}(m,m')
\end{equation}
for all $m,m' \in \caF$ and $t \geq s$.  In particular, quite
generally,
\begin{equation}\label{1p}
  H(\phi_t(m)) \leq H(\phi_s(m)) + J_{t,s}(\phi_t(m),\phi_s(m)), \qquad t \geq s
\end{equation}
while, as from \eqref{markov1}, $J_{t,s}(\phi_s(m),\phi_t(m))=0$.
On the other hand, if the dynamical system $(\Om^N,U^N_t,\rho^N)$
is reversible under an involution $\pi^N$, $U^N_t = \pi^N U^N_{-t}
\pi^N$ such that $\rho^N \pi^N = \rho^N$, $\pi^N m^N = m^N$, then
\begin{equation}\label{reversi}
  H(m') - J_{t,s}(m',m) = H(m) - J_{t,s}(m,m')
\end{equation}
for all $m,m',t\geq s$. Hence, under dynamical reversibility
\eqref{1p} is  an equality:
\begin{equation}\label{2p}
J_{t,s}(\phi_t(m),\phi_s(m))=  H(\phi_t(m)) - H(\phi_s(m)), \qquad
t \geq s
\end{equation}
Remarks on the \rm{H}-theorem for irreversible dynamical systems
have been written in \cite{nonl}.

\subsubsection{Propagation of constrained
equilibrium}\label{propa} The condition \eqref{markov1} of
autonomy needs to be checked for all times $t \geq s \geq 0$,
starting at time zero from an initial value $m$. Obviously, that
condition is somehow related to -- yet different from Boltzmann's
Stosszahlansatz. The latter indeed corresponds more to the
assumption that any initial constrained equilibrium state at time
zero evolves to new constrained equilibria at times $t > 0$.
Formally and in the notation of Section 2, we consider a region
$M_0$ in phase space corresponding to some macroscopic state and
its image $fM_0$ after some time $t$. We then have in mind to ask
that for ``relevant" phase space volumes $A$
 \begin{equation}\label{condition: constrained eq} \frac{|UM_0 \cap A|}{|UM_0|} =
\frac{|M_t \cap A|}{|M_t|}
\end{equation} which means that the
evolution takes the equilibrium constrained with $x\in M_0$ to a
new equilibrium at time $t$ constrained at $x\in M_t$, \emph{as
far as the event $A$ is concerned}. Indeed, we expect that one
cannot distinguish $M_t$ from $UM_0$ by looking at macroscopic
variables. Hence the events $A$ should correspond to values of
these macroscopic variables ($A_1$ as defined below is an
example). Since we also expect that one cannot distinguish $M_t$
from $UM_0$ by studying the future evolution, $A$ can also
correspond to values of the macrovariables in the future ($A_2$ as
defined below is an example).  However, the states $M_t$ and
$UM_0$ can (in principle) be distinguished by looking at their
past macrotrajectory. Indeed, one does not expect
\eqref{condition: constrained eq} to hold for $A=UM_0$ since the
left-hand side is $1$ and the right-hand side (which is dominated by $|M_0|/|M_t|$) is typically exponentially small in $N$.\\
 To be more precise and turning back to the
present mathematical context, we consider the following condition:
\begin{equation}\label{ce}
  \lim_{\ep \downarrow 0} \lim_{N\uparrow\infty} \frac{1}{N} \log
  \frac{\rho^N\{x \in A^N_i \rel m^N(x) \stackrel{\ep}{=} \phi_t(m)\}}
  {\rho^N\{U^N_t x \in A^N_i \rel m^N(x) \stackrel{\ep}{=} m\}} = 0
\end{equation}
for all $m \in \caF$, $t \geq s \geq  0, i=1,2$, $A^N_1 \equiv
\{m^N(x)
\stackrel{\de}{=} \phi_t(m)\}$ and\\
 $A^N_2 \equiv \{m^N(U^N_t
(U^N_s)^{-1} x) \stackrel{\de}{=} \phi_t(m)\}$.\\
Arguably, \eqref{ce} is a (weak) version of  propagation of
constrained equilibrium. We check that it actually implies
condition \eqref{markov1}, and hence the \rm{H}-theorem. First,
choosing $A^N_1 = \{m^N(x) \stackrel{\de}{=} \phi_t(m)\}$,
\eqref{ce} yields
\[
  \lim_{\ep \downarrow 0} \lim_{N\uparrow\infty} \frac{1}{N}
  \log\rho^N\{m^N(U^N_t x) \stackrel{\de}{=} \phi_t(m) \rel m^N(x) \stackrel{\ep}{=} m\}
  = 0
\]
Second, using the invariance of $\rho^N$,
\begin{multline*}
  \rho^N\{m^N(U^N_t x) \stackrel{\de}{=} \phi_t(m) \rel m^N(U_s^N x) \stackrel{\ep}{=} \phi_s(m)\}
\\
  = \rho^N\{m^N(U^N_t x) \stackrel{\de}{=} \phi_t(m) \rel m^N(x) \stackrel{\ep}{=} m\}\,
  \frac{\rho^N\{m^N(U^N_t (U^N_s)^{-1} x) \stackrel{\de}{=} \phi_t(m) \rel m^N(x) \stackrel{\ep}{=} \phi_s(m)\}}
  {\rho^N\{m^N(U^N_t (U^N_s)^{-1} U^N_s x) \stackrel{\de}{=} \phi_t(m) \rel m^N(x) \stackrel{\ep}{=} m\}}
\end{multline*}
and by applying condition \eqref{ce} once more but now with\\
$A^N_2 = \{m^N(U^N_t (U^N_s)^{-1} x) \stackrel{\de}{=}
\phi_t(m)\}$ and taking the limits, we get \eqref{markov1}. (Note
that we have actually also used here  that $U^N_t$ is invertible
or at least that $U^N_t \circ (U_s)^{-1}$, $t \geq s$ is well
defined.)

\subsection{Finite-size formulation\label{findim}}
 As announced in the questions under Section \ref{que}, we are
 certainly most interested in the case of finite but very large $N$ and how
 the H-function can be defined along the microscopic trajectory.\\
Consider
\begin{equation}\label{finite entropy}
  H^{N,\ep}(m) = \frac{1}{N}  \log\rho^N\{m^N(x) \stackrel{\ep}{=} m\}
\end{equation}
for (macrostate) $m \in \caF$. For a microstate $x \in \Om^N$,
define
\begin{align*}
  \overline H^{N,\ep}(x) &= \sup_{m \stackrel{\ep}{=} m^N(x)} H^{N,\ep}(m)
\\
  \un H^{N,\ep}(x) &= \inf_{m \stackrel{\ep}{=} m^N(x)} H^{N,\ep}(m)
\end{align*}

Instead of the hypothesis \eqref{markov1} of macroscopic autonomy,
we assume here that there is a macroscopic dynamics $\phi_t,
\phi_0=$ Id, for which
\begin{equation}\label{markov2}
\lim_{\ep \downarrow 0} \lim_{N \uparrow \infty} \rho^N\{m^N(U^N_t
x) \stackrel{\de}{=} \phi_t(m) \rel m^N(x) \stackrel{\ep}{=} m\}=1
\end{equation}
and that
\begin{equation}\label{markov3}
\lim_{\ep \downarrow 0} \lim_{N \uparrow \infty} \frac 1{N} \log
\rho^N\{m^N(U^N_t x) \stackrel{\ep}{=} \phi_t(m) \rel m^N(U_s x)
\stackrel{\ep}{=} \phi_s(m)\}=0
\end{equation}
for all $m \in \mathcal{F}$, $\delta > 0$ and $0 \leq s \leq t$.
Condition \eqref{markov2} corresponds to the situation in
\eqref{markov1} but where there is a unique macroscopic
trajectory, which one observes typically. Now we have
\begin{proposition}\label{ja5}
 Assume \eqref{markov2}-\eqref{markov3}. Fix a finite sequence of times $0<
t_1<\ldots<t_K$. For all $ m \in \cal{F}$, there exists $\de_0 >0$
such that for all $\de \leq \de_0$,
\begin{equation}\label{hn}
\lim_{\ep \downarrow 0} \lim_{N \uparrow \infty}
  \rho^N\big[
  \overline H^{N,\de}(U_{t_j}^Nx) \geq \un H^{N,\de}(U_{t_{j-1}}^Nx)
  -\frac{1}{N},\,j = 1,\ldots,K
  \rel m^N(x) \stackrel{\ep}{=} m\big]=1
\end{equation}
\end{proposition}

\begin{proof}
Put
\begin{equation}
g^{N,\de,\ep}(s,t,m)   \equiv 1- \rho^N\{m^N(U^N_t x)
\stackrel{\de}{=} \phi_t(m) \rel m^N(U_s x) \stackrel{\ep}{=}
\phi_s(m)\}
\end{equation}
then
\begin{equation}\label{byandby}
  \rho^N\big[m^N(U_{t_j} x) \stackrel{\de}{=} \phi_{t_j}(m), j=1,\ldots,K \rel
  m^N(x) \stackrel{\ep}{=} m\big] \geq 1 - \sum_{j=1}^K g^{N,\de,\ep}(0,t_j,m)
\end{equation}

Whenever $m^N(U_{t} x) \stackrel{\de}{=} \phi_{t}(m)$, then
\[ \un
H^{N,\de}(U^N_t x) \leq H^{N,\de}(\phi_t(m)) \leq \overline
H^{N,\de}(U^N_t x)\]
  As a
consequence, \eqref{byandby} gives

\begin{eqnarray}\label{byandby2}
 && \rho^N\big[\un H^{N,\de}(U_{t_j} x) \leq H^{N,\de}(\phi_{t_j}(m))
\leq \overline H^{N,\de}(U_{t_j} x), j=1,\ldots,K \rel
  m^N(x) \stackrel{\ep}{=} m\big]\nonumber \\
  && \geq 1 - \sum_{j=1}^K g^{N,\de,\ep}(0,t_j,m)
\end{eqnarray}
The last term can be controlled via \eqref{markov2}.  On the other
hand, by the same bounds as in \eqref{bounds}, we have
\[
H^{N,\de}(\phi_{t_j}(m)) \geq H^{N,\de}(\phi_{t_{j-1}}(m))
+\frac{1}{N} \log[1 - g^{N,\de,\de}(t_{j-1},t_j,m)]
\]
The proof is now finished by using \eqref{markov3} and choosing
$\de_0$ such that for $\de \leq \de_0$ and for large enough $N$
(depending on $\de$).
\[
\min_{j=1}^K{ \big( \log[1 - g^{N,\de,\de}(t_{j-1},t_j,m)] \big)}
\geq -1
\]
\end{proof}

\section{Additional remarks}\label{adrem} The above remains
essentially unchanged for stochastic microscopic dynamics. Instead
of the dynamical system $(\Omega,U_t,\rho)$ one considers any
stationary process $(X^N_t)_{t\in \bbR_+}$ with the law
$\textbf{P}_N$; denote by $\rho^N$ the stationary measure. The
entropy is defined as in \eqref{conbe} but with respect to
$\rho^N$. The (weak) autonomy in the sense of (\ref{markov1}) is
then
\[
 \lim_{\ep \downarrow 0} \lim_{N\uparrow +\infty} \frac 1{N} \log\textbf{P}_N[
m^N(X^N_t) \stackrel{\de}{=} \phi_t(m) \big|  m^N(X_s^N)
\stackrel{\ep}{=} \phi_s(m) ] = 0
\]

On the other hand some essential changes are necessary when
dealing with quantum dynamics.  The main reason is that, before
the limit $N\uparrow +\infty$, macroscopic variables do not
commute so that a counting or large deviation type definition of
entropy is highly problematic.  We keep the solution for a future
publication.\\

While some hesitation or even just confusion of terminology and
concepts have remained, the physical arguments surrounding an
\rm{H}-theorem have been around for more than 100 years.  The main
idea, that deterministic autonomous equations give an
\rm{H}-theorem when combined with the Liouville theorem, is
correct but the addition of some mathematical specification helps
to clarify some points.
 In this paper, we have repeated the following points:
 \begin{enumerate}
 \item
 There is a difference between the Second Law of Thermodynamics when considering transformations between
 equilibrium states, and microscopic versions, also in nonequilibrium contexts,
  in which the Boltzmann entropy is evaluated and plays the role
 of an \rm{H}-function.
 \item  The autonomy of the macroscopic equations should be understood
as a semigroup property (first order differences in time) and it
is a weaker condition than the one of propagation of equilibrium.
Mostly, that autonomy only appears sharply in the limit of
infinite scale separation between the microscopic world and the
macroscopic behavior.  A specific limiting argument is therefore
required to combine it with Liouville's theorem about conservation
of phase space volume for finite systems.
\end{enumerate}

As a final comment, there remains the question how useful such an
analysis can be today. Mathematically, an \rm{H}-theorem is useful
in the sense of giving a Lyapunov function for a dynamical system,
to which we alluded in the introduction. Physically, an
\rm{H}-theorem gives an extension and microscopic derivation of
the Second Law of thermodynamics. One point which was however not
mentioned here before, was much emphasized in years that followed
Boltzmann's pioneering work, in particular by Albert Einstein. The
point is that one can usefully turn the logic around.  The
statistical definition of entropy starts from a specific choice of
microstates. If for that choice, the corresponding macroscopic
evolution is not satisfying an \rm{H}-theorem, then our picture of
the microstructure of the system is very much expected to be
inadequate. In other words, we can obtain information about the
microscopic structure and dynamics from the autonomous macroscopic
behavior. Then, instead of concentrating on the derivation of the
macroscopic evolutions with associated \rm{H}-theorem, we use the
phenomenology to discover crucial features about the microscopic
world.  That was already the strategy of Einstein in 1905 when he
formulated the photon-hypothesis.


\bibliographystyle{plain}

\end{document}